\documentclass{article}
\usepackage{frascatiphys,here,graphicx,subfigure}
\begin{document}
\title{ 
Purely Flavored Leptogenesis at the TeV Scale
}
\author{
Luis Alfredo Mu\~noz\\
{\em Instituto de F\'isica, Universidad de Antioquia, A.A. 1226, Medell\'in, Colombia} \\
}
\maketitle
\baselineskip=11.6pt
\begin{abstract}
I study variations of the standard leptogenesis scenario that can arise if an additional mass scale related to 
the breaking of some new symmetry is present below the mass $M_{N_1}$ of the lightest right-handed Majorana neutrino. 
I present a particular realization of this scheme that allows for leptogenesis at the TeV scale. In this realization 
the baryon asymmetry is exclusively due to flavor effects.
\end{abstract}
\baselineskip=14pt
\section{Introduction}
\label{sec:int}
From observations of light element abundances and of the Cosmic
microwave background radiation \cite{Hinshaw:2008kr} the Cosmic 
baryon asymmetry, $Y_B=\frac{n_{B}-n_{\bar B}}{s}=(8.75\pm 0.23)\times 10^{-10}$, 
(where $s$ is the entropy density)
can be inferred. The conditions for a dynamical generation of this
asymmetry (baryogenesis) are well known \cite{Sakharov:1967dj} and
depending on how they are realized different scenarios for
baryogenesis can be defined (see ref. \cite{Dolgov:1991fr} for a
througout discussion).
Leptogenesis \cite{Fukugita:1986hr} is a scenario in which an initial
lepton asymmetry, generated in the out-of-equilibrium decays of heavy
singlet Majorana neutrinos ($N_\alpha$), is partially
converted in a baryon asymmetry by anomalous sphaleron
interactions~\cite{Kuzmin:1985mm} that are standard model processes.
Singlet Majorana neutrinos are an essential ingredient for the
generation of light neutrino masses through the seesaw mechanism
\cite{Minkowski:1977sc}.  This means
that if the seesaw is the source of neutrino masses then qualitatively,
leptogenesis is unavoidable.  Consequently, whether the baryon
asymmetry puzzle can be solved within this framework turn out to be a
quantitative question. This has triggered a great deal of interest on
quantitative analysis of the standard leptogenesis model and indeed a
lot of progress during the last years has been achieved (see
ref.\cite{Davidson:2008bu}). 
\section{The Model}
\label{sec:model}
The model we consider here \cite{AristizabalSierra:2007ur} is a simple
extension of the standard model containing a set of $SU(2)_L\times U(1)_Y$
fermionic singlets, namely three right-handed neutrinos ($N_\alpha = N_{\alpha
  R} + N_{\alpha R}^c$) and three heavy vectorlike fields ($F_a=F_{aL} +
F_{aR}$). In addition, we assume that at some high energy scale, taken to be
of the order of the leptogenesis scale $M_{N_1}$, an exact $U(1)_X$ gauge horizontal
symmetry forbids direct couplings of the lepton $\ell_i$ and Higgs $\Phi$
doublets to the heavy Majorana neutrinos $N_\alpha$.  At lower energies,
$U(1)_X$ gets  spontaneously broken by the vacuum expectation value (vev)
$\sigma$ of a $SU(2)$ singlet scalar field $S$.  Accordingly, the Yukawa
interactions of the high energy Lagrangian read
{
\begin{equation}
\label{eq:lag}
  -{\cal L}_Y = 
  \frac{1}{2}\bar{N}_{\alpha}M_{N_\alpha}N_{\alpha} +
  \bar{F}_{a}M_{F_a}F_{a} +
  h_{ia}\bar{\ell}_{i}P_{R}F_{a}\Phi + 
\bar{N}_{\alpha}
\left(  \lambda_{\alpha a} + \lambda^{(5)}_{\alpha a}\gamma_5\right)
     F_{a}S.
\end{equation}
}
We use Greek indices $\alpha,\beta\dots =1,2,3$ to label the heavy Majorana
neutrinos, Latin indices $a,b\dots =1,2,3$ for the vectorlike messengers, and
$i, j, k, \dots$ for the lepton flavors $e,\mu,\tau$.  Following reference
\cite{AristizabalSierra:2007ur} we chose the simple $U(1)_X$ charge
assignments $X(\ell_{L_i},F_{L_a},F_{R_a})=+1$, $X(S)=-1$ and
$X(N_{\alpha},\Phi)=0$.  This assignment is sufficient to enforce the absence
of $\bar N \ell \Phi$ terms, but clearly it does not constitute an attempt to
reproduce the fermion mass pattern, and accordingly we will also avoid
assigning specific charges to the right-handed leptons and quark fields that
have no relevance for our analysis.  
As discussed in~\cite{AristizabalSierra:2007ur}, depending on the
hierarchy between the relevant scales of the model
($M_{N_1},\,M_{F_a},\,\sigma$), quite different {\it scenarios} for
leptogenesis can arise: $i$) For $M_F,\sigma\gg M_N$, we recover the Standard 
Leptogenesis (SL) case that will not discuss. 
$ii$) For $\sigma < M_{N_1}< M_{F_a}$ we obtain the Purely Flavored Leptogenesis 
(PFL) case, that corresponds to the situation, when the flavor symmetry $U(1)_X$ is 
still unbroken during the leptogenesis era and at the same time the messengers 
$F_a$ are too heavy to be produced in $N_1$ decays and scatterings, and can be 
integrated away (for other possibilities see ref.~\cite{AristizabalSierra:2007ur}).
After $U(1)_X$ and electroweak symmetry breaking
the set of Yukawa interactions in (\ref{eq:lag}) generates light
neutrino masses through the effective mass operator. The resulting mass matrix can be written as
\cite{AristizabalSierra:2007ur}
\begin{equation}
  \label{eq:nmm}
  -{\cal M}_{ij}=
  \left[
    h^{*}\frac{\sigma}{M_{F}}\lambda^{T}\frac{v^{2}}
    {M_{N}}\lambda\frac{\sigma}{M_{F}}h^{\dagger}
  \right]_{ij}
  = \left[
    \tilde{\lambda}^{T}\frac{v^{2}}{M_{N}}\tilde{\lambda}
  \right]_{ij} \,.
\end{equation}
Here we have introduced the seesaw-like couplings
  $\label{eq:seesaw-couplings}
  \tilde{\lambda}_{\alpha i} = 
  \left(
    \lambda \frac{\sigma}{M_F}h^\dagger
  \right)_{\alpha i}\,$.
Note that, in contrast to the standard seesaw, the
neutrino mass matrix is of fourth order in the {\it fundamental}
Yukawa couplings ($h$ and $\lambda$) and due to the factor
$\sigma^2/M_F^2$ is even more suppressed.
\section{Purely flavored leptogenesis}
\label{sec:pfl}
\begin{figure}[!ht]
  \begin{center}
    {\includegraphics[scale=0.5]{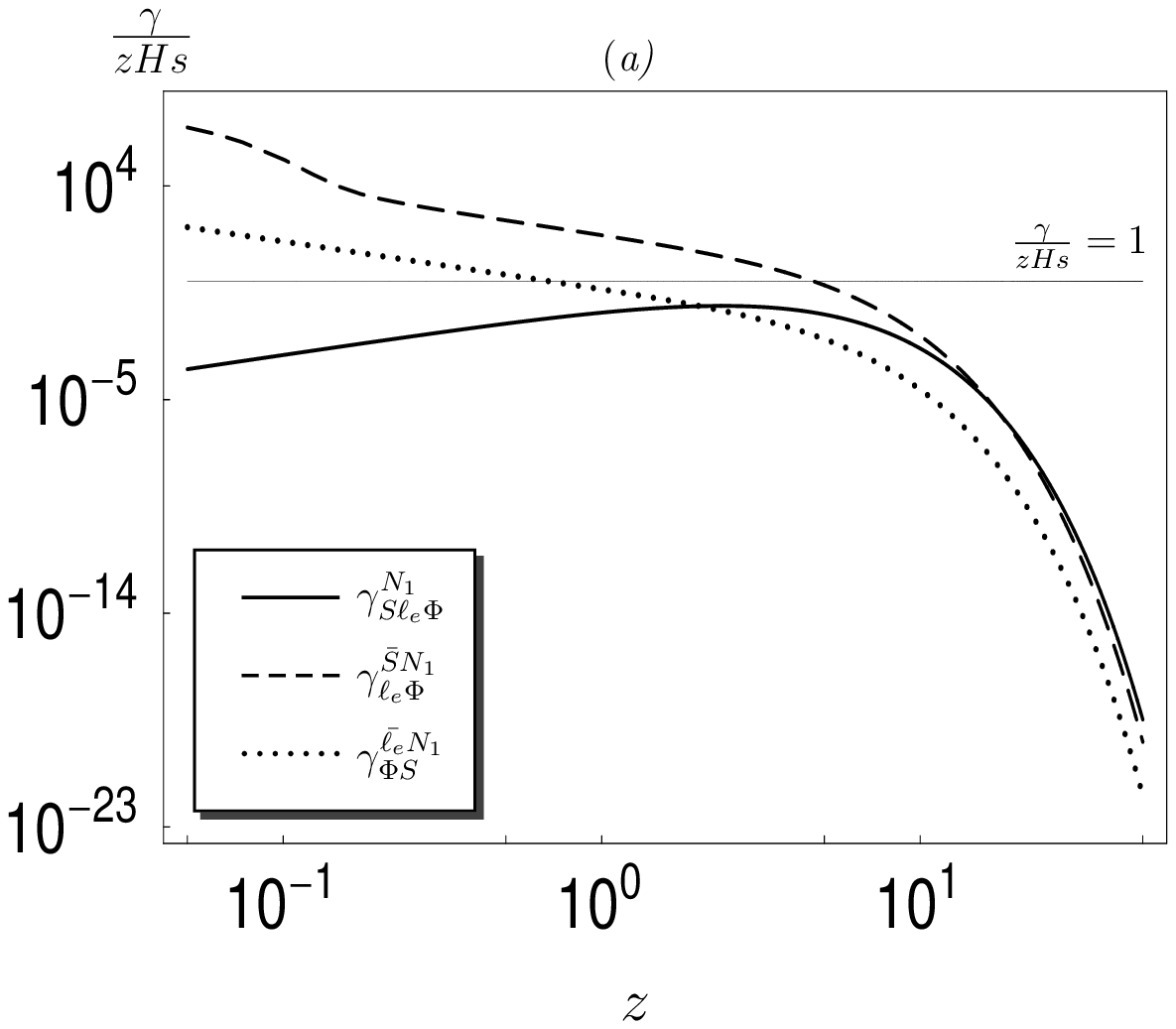}}
    {\includegraphics[scale=0.5]{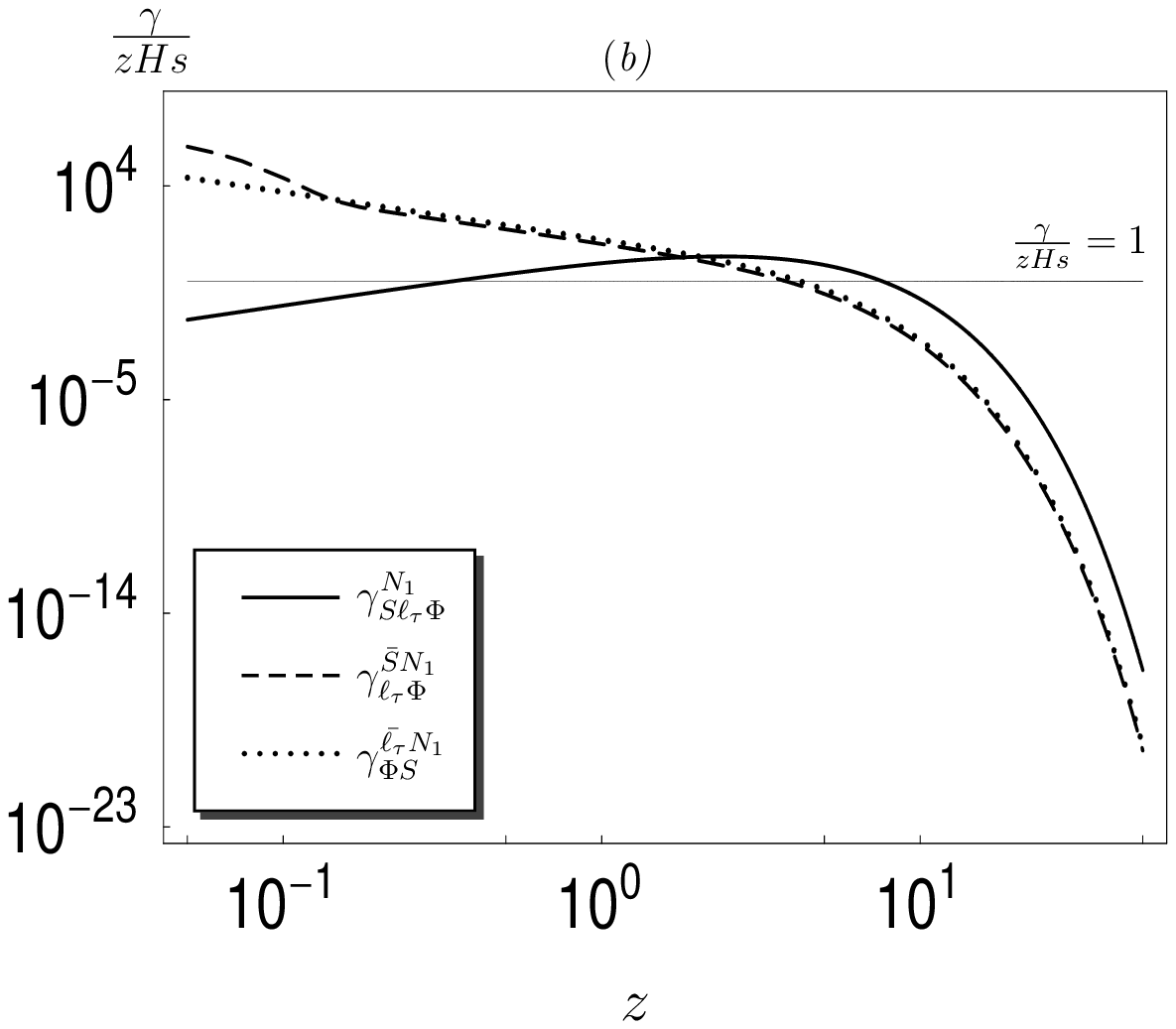}}
      \caption{({\it a}) The $e$ density rates. Around $z\sim1$ {\it s}-channel scatterings is the dominant one.
      ({\it $b$}) The $\tau$ density rates. Around $z\sim1$ both decay and scatterings are faster than $z\, H\, s$.}
\label{rates-el-tau-fig}
    \end{center}
\end{figure}

\begin{figure}
    \begin{center}
        {\includegraphics[scale=0.5]{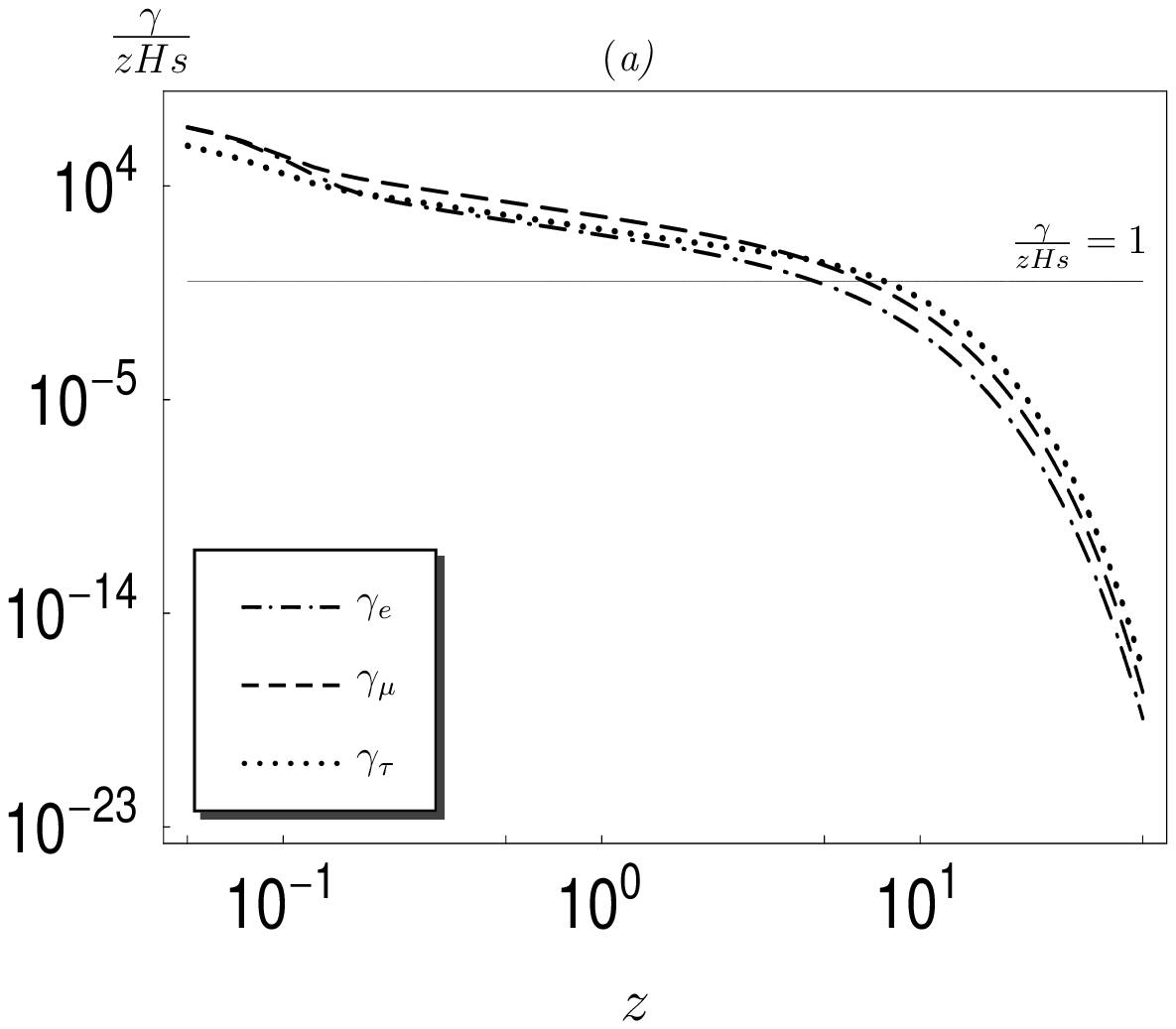}}
        {\includegraphics[scale=0.5]{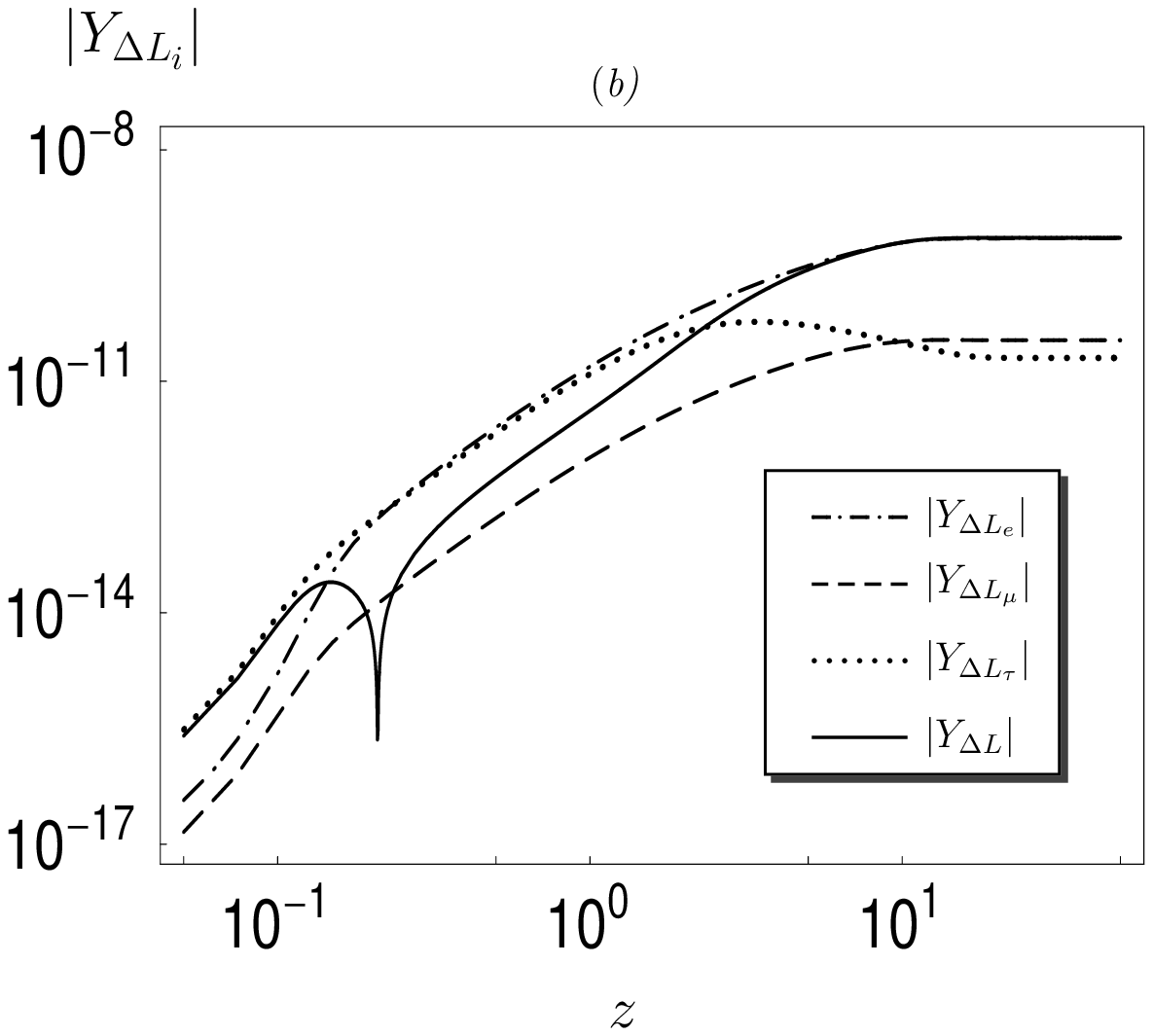}}
        \caption{({\it a}) Washout rates: with our choice of parameters the washout for the $\tau$ is 
	stronger (weaker) than that of $e$ and $\mu$, for $z\gg1$ ($z\ll1$). ({\it b}) Evolution of the 
	flavored and of the total lepton asymmetry. The final value is $Y_{\Delta L}=-7.1\times10^{-10}$,
	that is approximately three times larger than the experimental value, and is completely dominated 
	by $Y_{\Delta L_{e}}$.}
\label{rates-yl-fig}
    \end{center}
\end{figure}

In the case when $\sigma<M_{N_1}<M_{F}$, two-body $N_1$ decays are kinematically forbidden.  
However, via off-shell exchange of the heavy $F_a$ fields, $N_1$ can decay to the three
body final states $S\Phi l$ and $\bar S\bar\Phi \bar l$. 
The CP asymmetry is obtained from the interference between the tree level and loop diagrams; 
and reads \cite{AristizabalSierra:2007ur}:
\begin{equation}
\label{eq:CP}
\epsilon_{N_{1}\rightarrow\ell_{j}}\equiv\epsilon_{j}=\frac{3}{128\pi}\frac{\sum_{i}\Im 
m \left[\left(h r^{2} h^{\dagger}\right)_{ij}\tilde{\lambda}_{1i}\tilde{\lambda}^{*}_{1j} 
\right]}{\left(\tilde{\lambda}\tilde{\lambda}^{\dagger}\right)_{11}},
\end{equation}
where $r=M_{N_1}/M_{F}$. The CP asymmetries in (\ref{eq:CP}) have the following properties: 
$i)$ $\epsilon_{j}\neq0$. 
$ii)$ The total CP asymmetry $\epsilon_{N_1}=\sum_{j}\epsilon_{j}=0$. This is because 
$\Im m [\tilde{\lambda}hr^{2}h^{\dagger}\tilde{\lambda}^{\dagger}]_{11}=0$, 
and is related to the fact that the loop does no involve lepton number violation.
$iii)$ Rescaling the couplings $h$ and $\lambda$ by a parameter $\kappa>1$ according to: 
$h\rightarrow \kappa h,\hspace{0.3cm}\lambda\rightarrow \kappa^{-1} \lambda$; 
enhances the CP asymmetries as $\epsilon_{i}\rightarrow\kappa^{2}\epsilon_{i}$ \cite{Sierra:2009bh}. 
\begin{figure}
    \begin{center}
        {\includegraphics[scale=0.79]{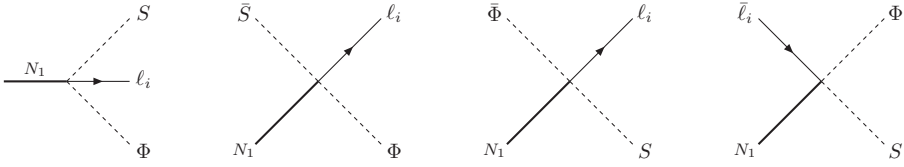}}
        \caption{ Relevant diagrams for leptogesis in our PFL model.}
\label{decay-scatter-fig}
    \end{center}
\end{figure}
In the Bolztmann Equations (BE), the particles densities are written in terms of the entropy density $s$, 
i.e $Y_{a}=n_{a}/s$ where $n_{a}$ is the number density for the particle $a$.  
We rescale the densities $Y_{a}$ by the equilibrium density $Y^{eq}_{a}$ of the corresponding particle, defining
$y_{a}\equiv Y_{a}/Y^{eq}_{a}$, while the asymmetries of the rescaled densities are denoted by 
$\Delta y_{a} \equiv y_{a}-y_{\bar{a}}$. In the BE, the time derivative is defined as 
$\dot{Y}_{a}=zHs\frac{d Y_{a}}{d z}$ where $z=M_{N_{1}}/T$ and $H(z)$ is the Hubble rate at temperature $T$. 
We denote the thermally averaged rate for an initial state $A$ to go the final state $B$ as 
$\gamma^{A}_{B}=\gamma\left(A\rightarrow B\right)$.
The processes in the BE are the decay of $N_1$ and the scatterings illustrated in figure \ref{decay-scatter-fig}; 
they are all of the same order ${\cal O}((\lambda^{\dagger} h)^2)$. The BE for the evolution of the $N_{1}$ abundance and 
of the lepton density asymmetry $Y_{\Delta L_{i}}$ are (see \cite{Sierra:2009bh} for details):
\begin{eqnarray}
  \label{eq:BE}
  \dot{Y}_{N_1}&=&-\left(y_{N_1}-1\right)\gamma\nonumber\\
  \dot{Y}_{\Delta L_{i}}&=& \left(y_{N_1}-1\right)\epsilon_{i}\gamma-
  \Delta y_{i}\left(\gamma_{i} +\left(y_{N_1}-1\right)\gamma^{N_{1}\bar\ell_{i}}_{S\Phi}\right),
\end{eqnarray}
where $\gamma_{i}=\gamma^{N_1}_{S\Phi\ell_i}+\gamma^{\bar{S}N_1}_{\Phi\ell_i}+
\gamma^{\bar{\Phi}N_1}_{S\ell_i}+\gamma^{\bar{\ell}_{i}N_1}_{S\Phi}$ is the sum of the
processes depicted en fig. \ref{decay-scatter-fig} and $\gamma=\sum_{i}\gamma_{i}+\bar{\gamma}_{i}$. 
For PFL the strong washout condition, corresponds to:$\left .\frac{\gamma}{z\,H\,s}\right|_{z\sim 1}>1$
where the normalization factor ${z\,H\,s}$ has been chosen to obtain an adimensional ratio. 
For the BE's solution we have chosen the couplings $h$ and $\lambda$ such that they reproduce the low 
energy neutrino parameters within $2\,\sigma$ and also satisfy the strong washout condition. The densities 
rates for $e$ and $\tau$ are shown in figure \ref{rates-el-tau-fig}, their expressions can be found in 
\cite{Sierra:2009bh}.
It is important to remark that the value of the lowest Majorana mass has be chosen 
at the TeV scale: $M_{N_1}=2.5$ TeV. Others parameters are $M_{N_2}=10$ TeV, $M_{N_3}=15$ TeV 
and  $r_{a}=[.1,.01,.001]$. The total reaction densities that determine the washout rates for the
different flavors are shown in the first panel in figure~\ref{rates-yl-fig}. 
Since PFL is defined by the condition that the sum of the flavor CP asymmetry vanishes 
($\sum_j \epsilon_{1j}=0$), it is the hierarchy between these washout rates that in the end is 
the responsible for generating a net lepton number asymmetry.  The total lepton asymmetry obtained 
is above the experimental data, but the couplings can be rescaled by a suitable value of $\kappa$ 
to get the right value.
\section*{Conclusions}
Variations of the standard leptogenesis scenario can arise from the
presence of an additional energy scale different from that of lepton
number violation. Quite generically, the resulting scenarios can be
expected to be qualitatively and quantitatively different from SL. 
Here we have considered what we regard as the simplest
possibility namely, the presence of an Abelian flavor symmetry
$U(1)_X$. The model allows for the possibility of generating the
Cosmic baryon asymmetry at a scale of a few TeVs. Moreover, our analysis
provides a concrete example of PFL, and shows that the condition
$\epsilon_1\neq 0$ is by no means required for successful leptogenesis.
\end{document}